\documentclass[twocolumn,amsmath,amssymb,pra,showpacs]{revtex4}

\usepackage[english]{babel}
\usepackage[applemac]{inputenc}
\usepackage[T1]{fontenc}
\usepackage{amsmath}
\usepackage{verbatim}
\usepackage[pdftex]{graphicx}
\usepackage[normalem]{ulem}
\usepackage{color}

\definecolor{dgreen}{rgb}{0.0, 0.7, 0.0}

\begin{document}

\bibliographystyle{aip}

\title{Off-resonant many-body quantum carpets in strongly tilted optical lattices}

\author{Manuel H. Mu\~noz-Arias, Javier Madro\~nero, and Carlos A. Parra-Murillo}
\affiliation{Departamento de f\'isica, Universidad Del Valle, A. A. 25360,Cali, Colombia\\
$^{*}$Correspondence should be addressed to \textbf{carlos.alberto.parra@correounivalle.edu.co}}

\begin{abstract}
An unit filling Bose-Hubbard Hamilonian embedded in strong Stark field is studied in the off-resonant 
regime inhibiting single and many-particle first order tunneling resonances. We investigate the occurrence 
of coherent dipole wave-like propagation along of an optical lattice by means of an effective Hamiltonian
accounting for second order tunneling processes. It is shown that dipole wavefunction evolution in the short-time
limit is ballistic and that finite size effects induce dynamical self-interference patterns known as 
quantum carperts. We also present the effects of the border right after the first reflection showing that
the wavefunction diffuses normally with the variance changing linearly in time. This work extends the rich 
physical phenomenology of the tilted one dimensional lattice systems in a scenario of many interacting 
quantum particles, the so-called many-body Wannier-Stark system.
\end{abstract}

\date{\today} 

\maketitle

\section{INTRODUCTION}
\label{sec:1}
The Bose-Hubbard Hamiltonian, a model that provides a good description of superconducting systems and the 
motion of strongly correlated quantum particles in a lattice system \cite{Knollman1963,Fisher1989,Zoller1998}, 
still remains a paradigmatic model in quantum mechanics. In one dimension, the single-band approach is the 
simplest version of this many-particle model. However, despite its simple form it contains subtle effective dynamics
relying on specific regimes of its parameter space \cite{Lewenstein2007,Bloch2008,Greiner2002}. 
This Hamiltonian not only allows for the construction of analog systems, or quantum simulators, for solid state 
systems as the Wannier-Stark system \cite{Salomon1996,Kolovsky2002,Simon2011,Meinert2013,Meinert2014,Sachdev2002,Kolovsky2004}, 
and its recent development including higher Bloch bands \cite{Sias2009,Plotz2010-11,Parra2013-14}, 
but also permits the characterization of quantum phase transitions that have been reported in experiments with ultracold atoms 
trapped in optical lattices \cite{Simon2011,Greiner2002,Meinert2013}. Within the context of many-body physics
in a lattice, the last decades have been very exciting after the seminal work of Jaksch {\it et al.} 
\cite{Zoller1998} and the development of experimental techniques for cooling, trapping and loading of cold atoms
in optical potentials \cite{Bloch2008}. This paved the way for the study of quantum matter and its interaction by means of 
magnetic fields that controls the scattering properties of colliding atoms, and therefore, their interaction strengths. Many-body scenarios 
beyond the mean field limit are thus nowadays at hand and captivate the attention of both experimentalists and theoreticians
due to its applicability in, for instance, quantum information processing \cite{Amico2008}. Here, detection techniques 
are extremely important for the control of systems at the level of single lattice sites \cite{Endres2013,Fukuhara2013,Greiner2015} 
permiting to study in a clean manner the evolution of single quasiparticles made of, for instance, two bound particles 
\cite{Simon2011,Meinert2013,Meinert2014} or unoccupied lattice sites which show to behave as quantum entities with a well 
defined dispersion relation \cite{Fukuhara2013}.

The Bose-Hubbard Hamiltonian is then one of the favorite toy models for studying quantum simulation and for testing
fundamental features of quantum mechanics. As example of this, physical phenomena as quantum magnetism in one-dimensional 
systems accounting also for superexchange \cite{Aidelsburger2011}, and even for the simulation of Higgs 
modes in optical lattices \cite{Kim2010} within the context of gauge fields, have been observed. Its realisation demands 
an overhelming control of experimental procedures specially in the many-body regime. Fortunately, the implementation of 
controlled dynamics by means of quantum quenches and sweeps has been amazingly improved allowing for the study of real 
time dynamical passages across critical points or regions characterized by strong spectral correlations, i.e.,  avoided crossings 
\cite{Simon2011,Meinert2013,Meinert2014,Arimondo2012-14,Chen2011,Polkovnikov2005,Wilkinson1988}.

In this paper, we focus our attention on the study of time evolution of strongly coupled quasiparticles so-called
doublon and holon. It is seen that, together, these quantum objects can move along the lattice as a unique 
quasiparticle, a dipole (or exciton), whose in-lattice motion is induced by a strong Stark field. There have been recent 
studies in that direction in which the focus was on the dynamics of a single holon (an empty 
lattice site) \cite{Andraschko2015}, or single double occupancy (doublon) \cite{Hofmann2012}, and or single 
impurities in a two-species untilted Bose-Hubbard model \cite{Fukuhara2013}. We show that the effects therein studied 
can be combined by applying a strong external Stark field to the lattice which also induces other interesting 
dynamical effects as quantum interference. Here, an initial state consisting of a single doublon-holon quasiparticle 
(a quantum dipole) in the tilted system experiences coherent evolution in time when the dipole moves along the lattice. The 
presence of boundaries implies a transit back and forth from which self-interference patterns 
arise, the so-called quantum carpets. These can be characterized by means of an effective Hamiltonian accounting for 
second order tunneling processes. Our analysis is restricted to the regime of strong Stark field $F$ such that $F \sim U$, 
with $U$ being the strength of the interparticle interaction. Of particular interest is the case for which 
the hopping amplitude $J$ is much smaller than the interaction strength $U$. This regime has been shown to allow 
for the appeareance of dynamical phenomena such as solitons \cite{Giorgios2011} and higher order tunneling 
processes with characteristic strength $J^m/U^{m-1}$ \cite{Meinert2014}, for $m$ an integer defining the order of 
the tunneling process.

This work is organized as follows: In section~\ref{sec:2}, we present the analysis of second order 
tunneling processes in the tilted single-band Bose-Hubbard Hamiltonian in the off-resonant regime.
In section~\ref{sec:3}, we construct an effective Hamiltonian intending to describe in a more accurate
way the in-lattice dipole propagation and the emergent self-interference patterns. In section~\ref{sec:4},
we study the short-time diffusive properties of the dipole wavefunction and the characterization of the
quantum carpets in terms of the size of the system and the boundary conditions. Section~\ref{sec:5}
summarizes our results and presents the conclusions.

\begin{figure}[b!]
\includegraphics[width=0.9\columnwidth]{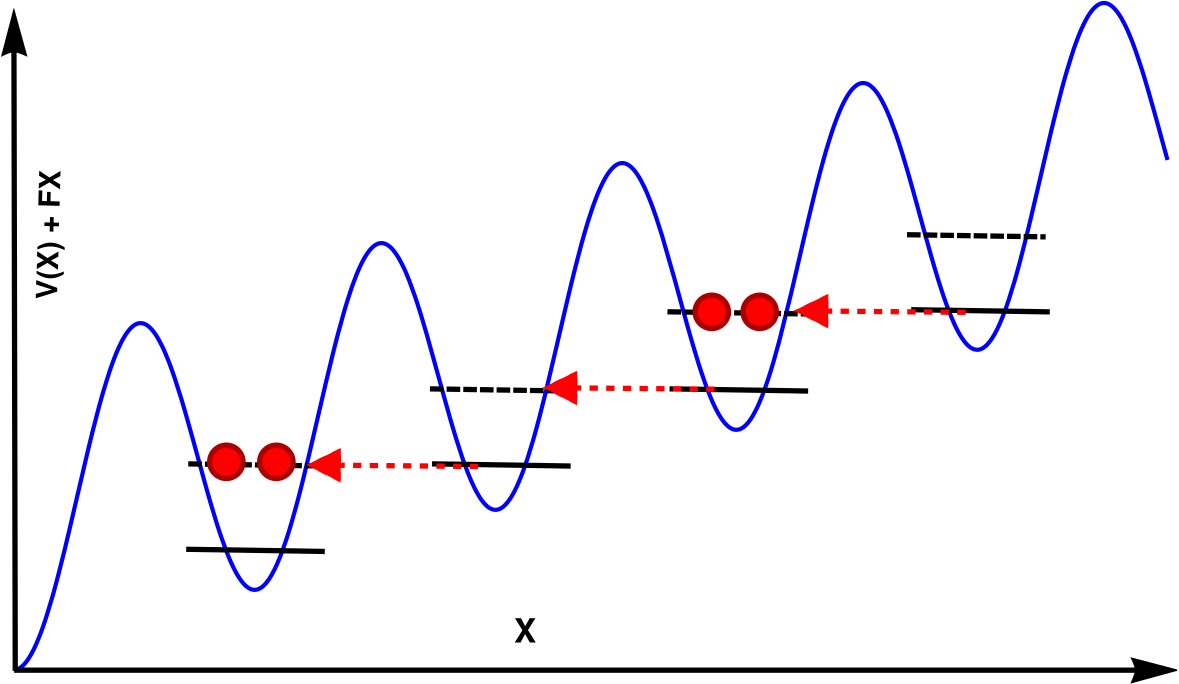}
\caption{\label{fig:1} (Color online) Schematic representation of doublon-holon 
creation in a strongly tilted optical lattice triggered via resonant tunneling between 
single (solid line) and two-particle (dashed line) states at $F\approx U$.}
\end{figure}

\section{SECOND ORDER PROCESS IN THE STRONG TILTING REGIME}
\label{sec:2}
The starting point in our analysis is the single-band one-dimensional Bose-Hubbard Hamiltonian 
for $N$ bosonic atoms within $L$ lattice sites \cite{Kolovsky2003}
\begin{equation}\label{eqn:1}
\hat H = \sum_lF l\hat n _l+\frac{U}{2}\hat n_l(\hat n_l-1)-J(\hat a_{l+1}^\dagger \hat a_l + {\rm h.c.})\,, 
\end{equation}
 Here we set $\hbar=1$ and the lattice constant is also set to one. The operators 
$\hat a_l^\dagger$ and $\hat a_l$ are the bosonic creation and annihilation operators, and $\hat n_l$ is the number 
operator. We assume unit filling lattice, that is $s_0\equiv N/L=1$, for which the dimension of the Fock basis, 
$\{|n_0n_1\cdots n_{L-1}\rangle\}$, is $d_{\rm Fock}=(2L-1)!/[L!(L-1)!]$. We restrict our analysis to the regime of strong
Stark field such that $F \sim U$, and $\{J,\,|U-F|\} \ll \{U,\,F\}$. As shown by Sachdev 
{\it et al.}~\cite{Sachdev2002}, within this regime first order processes allow for the creation of dipole excitation consisting 
of a pair doublon-holon, that is, a state with two particles in a well and an unoccupied the nearest 
neighboring well as sketched in Fig.~\ref{fig:1}. The type of states with only one doublon-holon, i.e., a state with 
the form $|111\cdots 201\cdots 1\rangle$ will be the main focus in our analysis.

To successfully generate appropriate conditions for the propagation of dipoles, a first step has to be done: 
To prepare of an initial state $|\psi_0\rangle=|111\cdots 201\cdots 1\rangle$. To do so, a Mott insulator is 
prepared in an untilted optical lattice ($F=0$) which is expected to be sufficiently stable in the chosen parameter 
regime (see \cite{Greiner2002}). At this stage a doubly occupied site plus an empty one can be created 
by means of single-site addressing techniques reported in~\cite{Fukuhara2013}. Given the previous 
step, our initial scenario is completed after suddendly quenching the Stark field from 
$\hat H(F=0)\rightarrow \hat H(F\neq 0)$ such that $F>0$ is far apart from any possible single and/or many-particle tunneling 
resonances, that is, $\lambda\equiv F/U>1$. Thereby, the system is expected to be dynamically frozen since 
first-order tunneling processes are not allowed by construction. The creation of a doublon implies the simultaneous 
creation of an empty site, here, reffered to as holon owing to its quasiparticle nature \cite{Andraschko2015}. We have then a 
quasiparticle consisting of a hole and a doublon, both together forming a quantum dipole. From the point of 
view of spin-1/2 systems this procedure corresponds to the flipping of only one spin in a chain~\cite{Fukuhara2013}.
Within this context it will be shown later that dipole propagation relates to the superexchange interaction. 

The set of states of the type $\{|111\cdots 201\cdots 1\rangle,\cdots\}$, i.e., those ones having a doubly occupied- and
an empty site togheter, spans an excitation manifold accessible by local hopping processes from the Mott-insulating 
state $|1111111\cdots 1\rangle$. This manifold is characterized by the conservation of the quantum dipole number
$\langle \sum\nolimits_l\hat d_l^{\dagger}\hat d_l\rangle_t$, where $\hat d_l^{\dagger}\propto\hat a^{\dagger}_{l-1}\hat a_l$ 
is the creation operator for dipoles. The dimension of this manifold, $d_{\rm dip}$, is very much smaller than the dimension, 
$d_{\rm Fock}$, of the Fock space corresponding of the full Bose-Hubbard Hamiltonian. In a previous work was shown that 
within this manifold subspace the Bose-Hubbard Hamiltonian can be mapped onto an antiferromagnetic spin-1/2 chain 
embedded in a transversal and longitudinal magnetic fields \cite{Sachdev2002}. This extraordinary simplification 
was experimentally confirmed by the group of Greiner {\it et al.} \cite{Simon2011}, thus establishing one of the first 
realization of a quantum simulation of one-dimensional quantum magnetism in optical lattices.
\begin{figure}[t!]
\includegraphics[width=0.95\columnwidth]{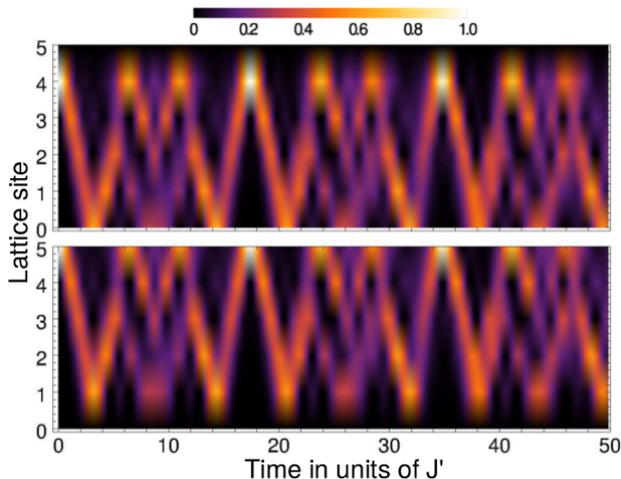}
\caption{\label{fig:2} (Color online) Density profile showing the propagation in time of (top) the doublon 
and (bottom) the  holon for $\lambda = 3$ which are constructed by the filtering procedure describes in the main text.
The initial state was $|111120\rangle$, for $L=6$.  Note that both doublon and holon follow the same dynamical
behavior along the lattice and are always one beside the other thus confirming their motion as that of a 
single unique quasiparticle, the doublon-holon or dipole excitation. The BH parameters are those mentioned 
in text and $J'$ is defined in Eq.~(\ref{eqn:6}).}
\end{figure}

We investigate dynamical effects beyond the resonant regime depicted in Fig.~\ref{fig:1}. To do so, we evolve 
in time the initial state $|\psi_0\rangle=|111\cdots 120\rangle$ and compute the lattice site doublon and holon
occupation numbers 
\begin{eqnarray}\label{eqn:2}
\langle \hat b_l\rangle_t&=&\frac{1}{2}\langle\psi_0|\hat U^{\dagger}_t\, [\hat n_l(\hat n_l-1)]\,\hat U_t|\psi_0\rangle\nonumber\\
\langle \hat h_l\rangle_t&=&\frac{1}{2}\langle\psi_0|\hat U^{\dagger}_t \,[(\hat n_l-1)(\hat n_l-2)]\,\hat U_t|\psi_0\rangle\,,
\end{eqnarray}
to track the dipole propagation along the lattice. The time evolution operator is defined by $\hat U_t=\exp(-i\hat H t)$. 
Note that the observables defined in Eq.~(\ref{eqn:2}) allows us to filter the individual dynamical evolution of
the doublon and the holon. However, as shown in Fig~\ref{fig:2}, it is now clear that these two move together 
without splitting across the lattice . This means that in order to characterize the dipole 
wavefunction it is enough to track one of its components, i.e., either the doublon or the holon. From Fig.~{\ref{fig:2}} 
we also recognize that the dipole wavefunction delocalizes in space in its in-lattice motion from one lattice edge to the opposite
one. The dipole wavefunction then starts highly localized and undergoes a spreading processes in its first edge-to-edge 
transit. This delocalization takes place as a smooth transition between manifold states, that is, from 
$|111\cdots 120\rangle \rightarrow |111\cdots 1201\rangle$ which is not possible by 
means of a single hopping action. At the opposite edge the dipole wavefunction is partially relocalized and the structure 
of the state is $|201\cdots 111\rangle$. This clean and coherent dynamics occurs within the $\lambda$-range 
$\{2.5,\cdots,4\}$, where $x\equiv F/U$. For $\lambda\gg 1$ the dipole remains frozen, i.e., it does not 
propagate along the lattice. The other limit, the one for which $\lambda\rightarrow 1$
implies an interplay between first- and second-order tunneling processes, which is translated to the creation/annihilation 
of dipoles induced by resonant tunneling effects at $F\approx U$, and the edge-to-edge dipole propagation, respectively.
From now onwards we fix $\lambda=3$ and the Bose-Hubbard parameters are taken from Ref.~\cite{Meinert2013}, that is 
$J=0.0254\,{\rm kHz}$ and $U=1.019\,{\rm kHz}$.

After the first edge-to-edge transit owing to the presence of boundaries the wavefunction gets reflected and the dipole is 
driven backwards. However, the subsequent transit is not as before since the wavefunction present a bifurcation 
which relates to dynamical self-interference effects. The mechanism for this to occur is the fact that both doublon and holon 
when reflected at the boundary adquire dynamical phases which influence the backward transit (see Fig.~\ref{fig:2}). 
Yet, irrespective of the dynamical self-interference it can be noticed that the localization properties of the dipole 
wavefunction are recovered and, naturally, after a certain time the initial state $|\psi_0\rangle$ is reconstructred 
resembling a sort of echo dynamics \cite{Hahn1950}. This localization-delocalization process is periodic in time and its 
characeristic period $T$ depends on the system size $L$.

In order to characterize this underlying dynamics of the strongly tilted Bose-Hubbard model in the off-resonant regime, we construct an 
effective Hamiltonian accounting for the main mechanism for the dipole propagation along the lattice. This effective model will 
allow us to describe accurately the physical phenomenology involved in the motion of 
quantum dipoles in an optical lattice. The major difficulty appearing when analyzing large systems in the full Bose-Hubbard 
context is that the dimension of the associated Fock space increases exponentially with the number of lattice sites. Then a 
systematic analysis of the above mentioned effects becomes untractrable for $L\gg 1$. With the effective model this study
gets easier.

\section{EFFECTIVE MODEL FOR DIPOLES}
\label{sec:3}
Since the Bose-Hubbard model~(\ref{eqn:1}) does not explicitly allow for the motion of particles beyond 
single hopping. We therefore construct an effective Hamiltonian accounting for two sucessive hopping processes
required for the dipole to move, that is, the transition between manifold states. This action requires
the participation of an auxiliary state, for example, the Mott-insulating state $|111111\cdots 1\rangle$. 
An efficient way to obtain the sort of model we look for is by means of the Schrieffer-Wolff 
transformation (SWT) \cite{Schrieffer1966,Chan2004}. The transformation consists in separating 
the Hamiltonian as $\hat H = \hat H_0 + \hat V$, where $\hat V$ is a perturbation 
to the Hamiltonian $\hat H_0$ for which its eigenvalues and eigenvectors are known. 
Since $J \ll U$ the hopping part might be treated as the perturbation $\hat V$. We now have to find out an 
antiunitary operator $\hat S$ such that
\begin{equation}\label{eqn:3}
\hat H^{\rm eff} = e^{\hat S} \hat H e^{-\hat S}\approx \hat H_0 + \frac{1}{2} [\hat S,\hat V]\,,
\end{equation}
given the constrain $\hat V = -[\hat S,\hat H_0]$. In general to find out $\hat S$ is not an easy task, but 
the use of the eigensystem solution of $\hat H_0$ allows us to write the matrix elements of $\hat H^{\rm eff}$ as
\begin{eqnarray}\label{eqn:4}
H^{\rm eff}_{ij} = \varepsilon_{i}\delta_{ij} + \frac{1}{2}\sum_{k\neq i,j} \left[\frac{1}{\varepsilon_i-\varepsilon_k}
-\frac{1}{\varepsilon_k-\varepsilon_j}\right]V_{ik}V_{kj}\,,
\end{eqnarray}
where $V_{ij}=\langle i|\hat V|j\rangle$. Note that the coupling between the eigenstates $|i\rangle$ and $|j\rangle$ 
of $\hat H_0$ via $\hat V$ involves the sum over all auxiliary states $|k\rangle\neq \,\{|i\rangle,|j\rangle\}$ 
for the required transition to occur. 
 
\begin{figure}[t!]
\includegraphics[width=0.95\columnwidth]{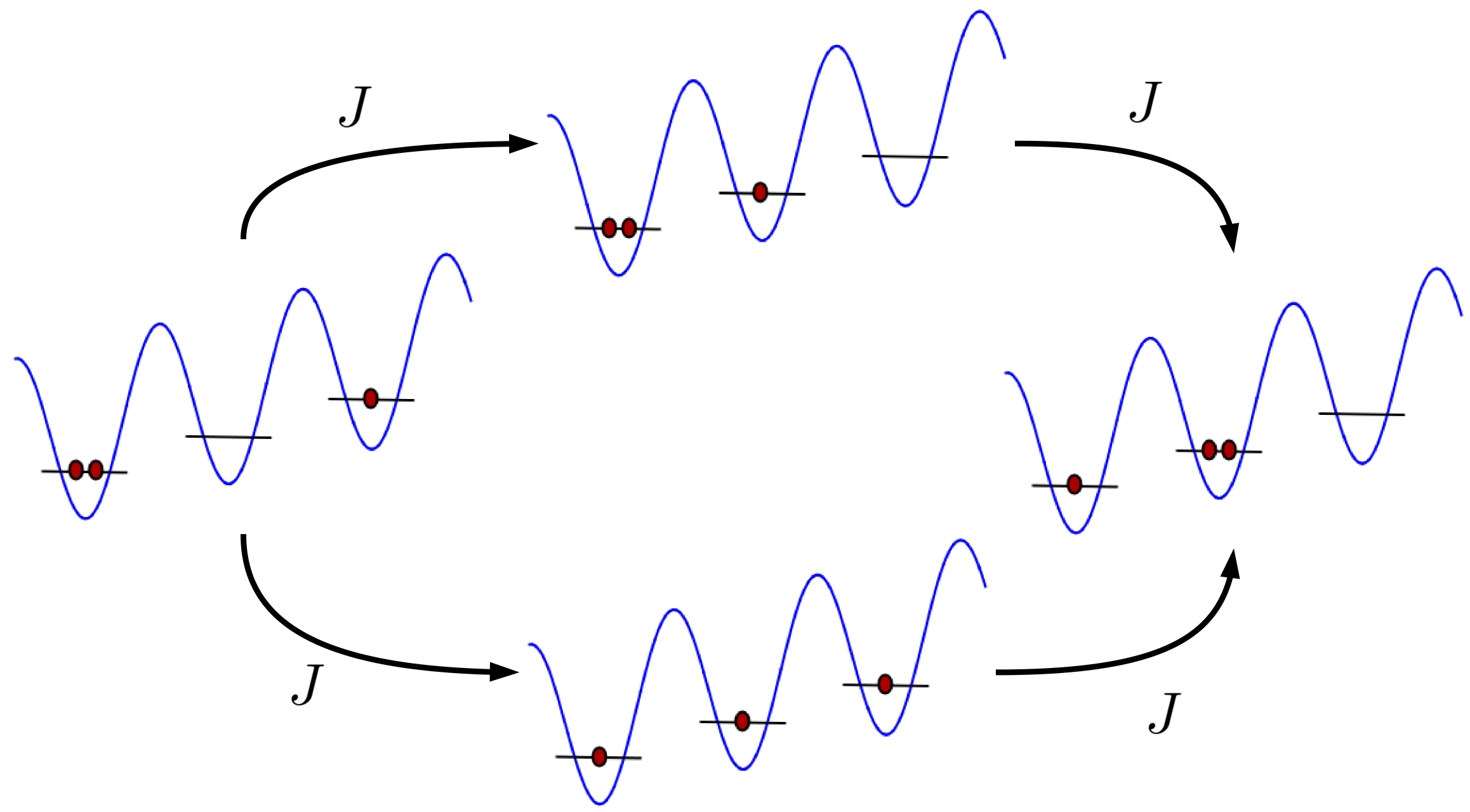}
\caption{\label{fig:3}(Color online) Schematic representation of the second order processes driving the transition 
between dipole states via two successive single-particle hopping processes involving auxiliary states.}
\end{figure}

We illustrate the implementation of Eq.~(\ref{eqn:4}) using the minimal system that can support dipole propagation, 
the system $N/L=3/3$. There are only two auxiliary states for the dipole hopping to occur (see fig.~\ref{fig:3}). First, 
to go from $|120\rangle$ to $|201\rangle$ an auxiliary state is the dipole vacuum given by the Mott state $|111\rangle$. 
Second, that transition is also possible via a second auxiliary state involving a holon and a doublon separated by one 
lattice site, that is, $|210\rangle$. The operator accounting for both of virtual processes is the same and its form is 
$a^\dagger_2 a_1a^\dagger_2a_3$. An additional reduction can be done on the final Hamiltonian $\hat H^{\rm eff}$ 
that will depend on the tilting term $\sim F l n_l$ of $\hat H_0$. We can get rid of it working in the interaction 
picture computed with respect to the term $\sum\nolimits_lFl\hat n_l$. This can be proven by showing 
that the commutator $[\hat H^{\rm eff}, \sum_l F l \hat n_l]=0$. We finally arrive to a 
translationally invariant and time-independent effective Hamiltonian for the dipoles given by
\begin{eqnarray}\label{eqn:5}
\hat H^{\rm eff} = \sum_l\frac{U}{2} \hat{n}_l (\hat{n}_l - 1) + 
J'  (\hat{a}^\dagger_{l+1}\hat{a}_l  \hat{a}^\dagger_{l+1}\hat{a}_{l+2} + {\rm h.c.})\,,
\end{eqnarray}
where
\begin{eqnarray}\label{eqn:6}
J'=2\frac{J^2}{U}\left(\frac{1}{\lambda}+\frac{1}{\lambda-1}\right).
\end{eqnarray}

\begin{figure}[b!]
\includegraphics[width=0.95\columnwidth]{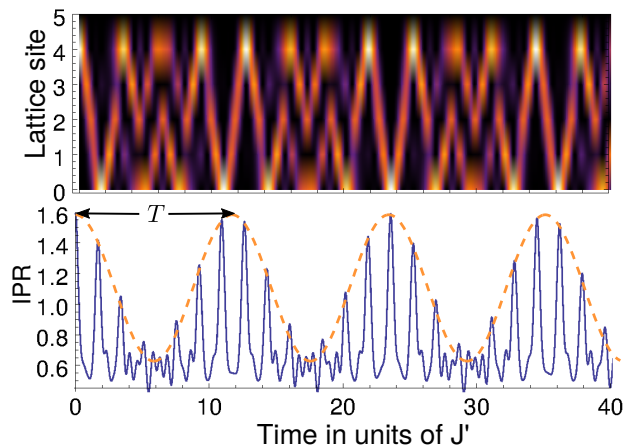}
\caption{\label{fig:4} (Color online) (Top) Density plot showing the evolution in time of the dipole excitation
(only doublon shown) which propagates along the lattice. This was computed using the effective Hamiltonian~(\ref{eqn:5}), 
with the initial state $|111120\rangle$ with open boundary conditions. (Bottom) IPR of the density plot as a function of time. 
The orange-dashed line represents the enveloping function that permits to recognize the periodical motion.}
\end{figure}

Note that now it is easy to recognize the dipole hopping term with strength $J'$ responsible for 
the dipole propagation along the lattice. In Fig.~\ref{fig:4}-(top) we present the density profile of
the doublon given the initial state $|\psi_0\rangle=|111120\rangle$. It was computed using the effective 
Hamiltonian~(\ref{eqn:5}) with open boundary conditions. As it can be seen, the effective model captures 
the principal characteristics of the full model~(\ref{eqn:1}), namely, propagation and dynamical 
interference (see Fig.~\ref{fig:4}-(top)).

Yet, there are subtle differences. For instance, in the framework of the effective Hamiltonian 
the dipole wavefunction after the first interference passage has to bounce twice at the edges in order to gain the 
dynamical phase enough to, again, trigger the self-interference pattern. As a consequence, the second pattern is 
inverted with respect to the first one. In spite of this dephasing-induced effect, one can notice that the time 
required for the appearance of this second pattern is the same in both full and effective model. 
 
Our dipole is a whole complex structure robust to the in-lattice propagation. This can be shown 
when mapping the bosonic hopping operators into dipole creation (annihilation) operators, as introduced in ref.~\cite{Sachdev2002},
\begin{equation}\label{eqn:7}
\hat d_l^\dagger = \frac{\hat a_{l+1}^\dagger \hat a_l}{\sqrt{s_0(s_0+1)}}\,,
\end{equation}
with $s_0=1$ the average lattice filling. The number of dipoles per site is restricted to either zero or 
one, and dipoles in nearest neighboring sites are not allowed. These restrictions translate to
\begin{equation}\label{eqn:8}
\hat d_l^{\dagger} \hat d_l \le 1 \qquad\text{and}\qquad \hat d_{l+1}^\dagger \hat d_{l+1} \hat d_l^\dagger \hat d_l =0\,.
\end{equation}
The effective model~(\ref{eqn:5}) under these constrains reduces to
\begin{equation}\label{eqn:9}
\hat H_{\rm dip} = \sum_l (U - F)\hat{d}_l^\dagger \hat{d}_l + 2J' (\hat{d}_{l+1}^\dagger \hat{d}_l + {\rm h.c.})\,,
\end{equation}
which describes the dipole as a single quasiparticle moving along the lattice. The term $U-F$ accounts for 
the detunning between the single and two-particle levels that allow for the creation of a doublon sketched in
Fig.~\ref{fig:1}. In our regime of interest this diagonal term can be neglected because of the dipole number
conservation.

The dipole representation is useful for the study of the localization properties of the dipole wavefunction.
This is done by defining the Inverse Participation Ratio (IPR)
\begin{equation}\label{eqn:10}
{\rm IPR} = \int_{\rm Lattice} |\rho(x,t)|^2 {\rm d}x\,,
\end{equation}
where $\rho(x,t)$ is the normalized dipole density probability distribution 
\begin{equation}\label{eqn:11}
\rho(x,t) = \sum_l \langle\hat d_l^\dagger\hat d_l\rangle_t\,\delta(x-l)\,,
\end{equation}
and we have defined
\begin{equation}\label{eqn:12}
\langle\hat d_l^\dagger\hat d_l\rangle_t=\langle\psi_0|e^{i\hat H_{\rm dip}t}\,\hat d_l^\dagger
\hat d_l\, e^{-i\hat H_{\rm dip}t}|\psi_0\rangle.
\end{equation}
In Fig.~\ref{fig:4}-(bottom) we plot the IPR correponding to the density profile of the system $L=6$. The IPR 
shows that indeed the dipole wavefunction delocalizes in every edge-to-edge transit where only two manifold
states are relevant, then IPR$_{\rm min}\approx 0.5$ (see Fig.~\ref{fig:4}-(bottom)). The self-interference 
regions are characterized by the spreading of the wavefunction, follwed by relocalization of this after certain 
time. This periodic behavior is highlighted by the enveloping function (orange-dashed line) from which the 
period $T$ can be extracted.

The propagation of dipoles can be interpreted even in a more fashionable representation consisting of spins.
Following Ref.~\cite{Sachdev2002} we use the pseudo spins transformation defined by $\hat\sigma_l^+ = 2 \hat d_l$, 
$\hat\sigma_l^- = 2 \hat d_l^\dagger$ and $\hat \sigma_l^z =1-2\hat d_l^{\dagger}\hat d_l$, to rewrite Eq.~(\ref{eqn:9})
in the spin representation resulting in the Heisenberg XXZ model
\begin{eqnarray}\label{eqn:13}
\hat H_{\rm spin} &= &\frac{W}{4} \sum_l \hat\sigma_l^z\hat\sigma_{l+1}^z  
+ J' \sum_l (\hat\sigma_l^x\hat\sigma_{l+1}^x   +  \hat\sigma_l^y\hat\sigma_{l+1}^y) \nonumber\\
&-& \frac{1}{2}(W+U-F) \sum_l \hat\sigma_l^z\,,
\end{eqnarray}
where $\hat\sigma_l^{x,y,z}$ are Pauli matrices, and the parameter $W$ is an extra energy term of order $U$. 
Here, it is clearly seen that the dipole propagation mechanism is nothing but the superexchange interaction which locally flips
two nearest neighboring spins in a chain. Then, it allows for the motion of only one flipped spin or spin impurity \cite{Fukuhara2013}
in a medium consisting of spins with opposite polarization. Likewise, the creation of a dipole translates to flipping 
one single spin in a chain with the Mott-insulating phase, $|111111\cdots 1\rangle$, represented by the state 
$|\downarrow\downarrow\downarrow\cdots\downarrow\rangle$. The strength of the effective magnetic field 
is $h =(W+U-F)/2$, and the anisotropy parameter is given by $\Delta = W/4J_{\rm exc}$, with $J_{\rm exc}=2 J'$ 
the superexchange coupling strength. Given the order of magnitude of the Bose-Hubbard parameters used
throughout this paper we have $\Delta\gg 1$ meaning that the dynamics of our system is deep inside 
the N\'eel (antiferromagnetic) phase of the Heisenberg XXZ model (see details in Ref.~\cite{Mikeska2004}). 

Unfortunately, the one dimensional XXZ model~(\ref{eqn:13}) is analitically untractable, that is, to obtain 
its eingenstates one has to implement numerical diagonalization, for instance, using the Lanczos algorithm \cite{ParraMurilloCPC}. 
Nevertheless, in our analysis we found that the effective dimensions of the new Hamiltonians (\ref{eqn:9}) and (\ref{eqn:13}) 
are much smaller than the dimension of the full Hamiltonian. In the case of the spin Hamiltonian the dimension 
is $d_{\rm spin}=2^L\ll d_{\rm Fock}$, and for the case of the dipole basis $\{|000\cdots 01_i0\cdots 0\rangle\}$ 
the number of accessible states $d_{\rm dip}=L\lll d_{\rm Fock}$. All this implies a fantastic reduction of 
computational resources for the analysis of suficiently large lattices. We then devote the rest of this
paper to study and characterize the self-interference effect and the diffusive behavior of our system 
usin these effective models.

\section{QUANTUM DIFFUSION AND EMERGENCE OF MATTER QUANTUM CARPETS}
\label{sec:4}
\subsection{Short-time dynamics}
\begin{figure}[t!]
\centering{\includegraphics[width=0.9\columnwidth]{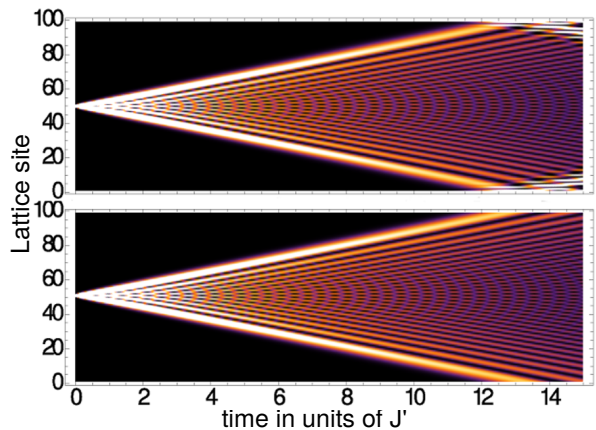}}
\caption{\label{fig:5}(Color online) Short-time evolution of the occupation site distribution for
$L=100$ with the dipole initially placed in the middle of the lattice. (Top) 
Using the effective Hamiltonian (\ref{eqn:9}) with open boundary conditions, and (bottom) 
Using the analytic expression~(\ref{eqn:15}).}
\end{figure}

The localization properties previously described can be studied in a more accurate way for the short-time
dynamics using the effective dipole Hamiltonian~(\ref{eqn:8}). Here, we show that the dipole wavefunction 
in the first edge-to-edge transit undergoes a ballistic delocalization, followed by a normally diffusive 
evolution right after its first reflection at the lattice borders. To do so, we assume open boundary conditions
and invoke the fact that the Hamiltonian $\hat H_{\rm dip}$ describing a quasiparticle confined in an untilted 
periodic potential is diagonal in the quasimomentum space. This latter is straighforwardly proven by expanding the bosonic lattice 
operators in their respective fourier series as $\hat d_l = L^{-1/2}\sum\nolimits_k e^{ikl} \hat d_k$, $\hat H_{\rm dip}$ 
transforms the effective Hamiltonian into 
\begin{eqnarray}\label{eqn:14}
 \hat H_{\rm dip}(k) = \sum_k \varepsilon(k) \hat d_k^\dagger \hat d_k\,,
\end{eqnarray}
with $\varepsilon(k) = F-U + 4 J' \cos(k)$ being the quasiparticle dispersion relation valid for suffiently large 
lattice. This formula brings advantages at obtaining an analitycal expression for the dipole site occupation 
$\langle\hat d_l^\dagger\hat d_l\rangle_t$~\cite{Hofmann2012} which results in
\begin{equation}\label{eqn:15}
 \langle\hat d_l^\dagger \hat d_l\rangle_t = \mathcal J_{l-l_0}^2 (4J' t)\,.
\end{equation}
$\mathcal J_{l-l_0}(t)$ is the Bessel function of the first kind and $l_0$ is the starting location of the dipole. 
The comparison of both numerical and theoretical density profiles is shown in Fig.~\ref{fig:5}, where not relevant 
differences are observed before the reflection at the edges. The ballistic wavefunction spreading in time is shown by 
computing the variance of the occupation dipole distribution~(\ref{eqn:10}) 
$\sigma^2=\langle \hat x^2\rangle-\langle \hat x\rangle^2$
for which an analytic expression is obtained using Eq.~(\ref{eqn:15}). The calculation follows as 
\begin{eqnarray}\label{eqn:16}
  \sigma^2&=&\sum_l l^2\mathcal J_{l-l_0}^2 (4J't)\left(1-\mathcal J_{l-l_0}^2 (4J't)\right)\nonumber\\
  &\approx&\sum_{l\neq l_0} l^2\mathcal J_{l-l_0}^2 (4J't)\,,
\end{eqnarray}
where we have used the fact that the Bessel function of the first kind, of the order $|l-l_0|>0$, are 
negligible if $4J' t\ll 1$. Now by expanding the Bessel function $\mathcal J_{l-l_0} (4J' t)$ up to the
first order we can write 
\begin{eqnarray}\label{eqn:17}
\mathcal J_{l-l_0}^2 (4J' t)&\approx& \frac{1}{(l-l_0)!^2}(2 J' t)^{2(l-l_0)}\nonumber\\
&+&\frac{1}{(l-l_0+1)!^2}(2J' t)^{2(l-l_0+2)}\nonumber\\
&-&\frac{2}{(l-l_0)!(l-l_0+1)!}(2J' t)^{2(l-l_0+1)}\,,\,\,\,\,\,\,\,
\end{eqnarray}
which helps us to evaluate Eq~(\ref{eqn:16}) to obtain the variance as a function
of time 
\begin{eqnarray}\label{eqn:18}
\sigma^2\approx\sum_{l\neq l_0} l^2\mathcal J_{l-l_0}^2 (4J't)=8(J't)^2\propto {\rm IPR}^{-1}\,.
\end{eqnarray}
\begin{figure}[t!]
\centering{\includegraphics[width=0.9\columnwidth]{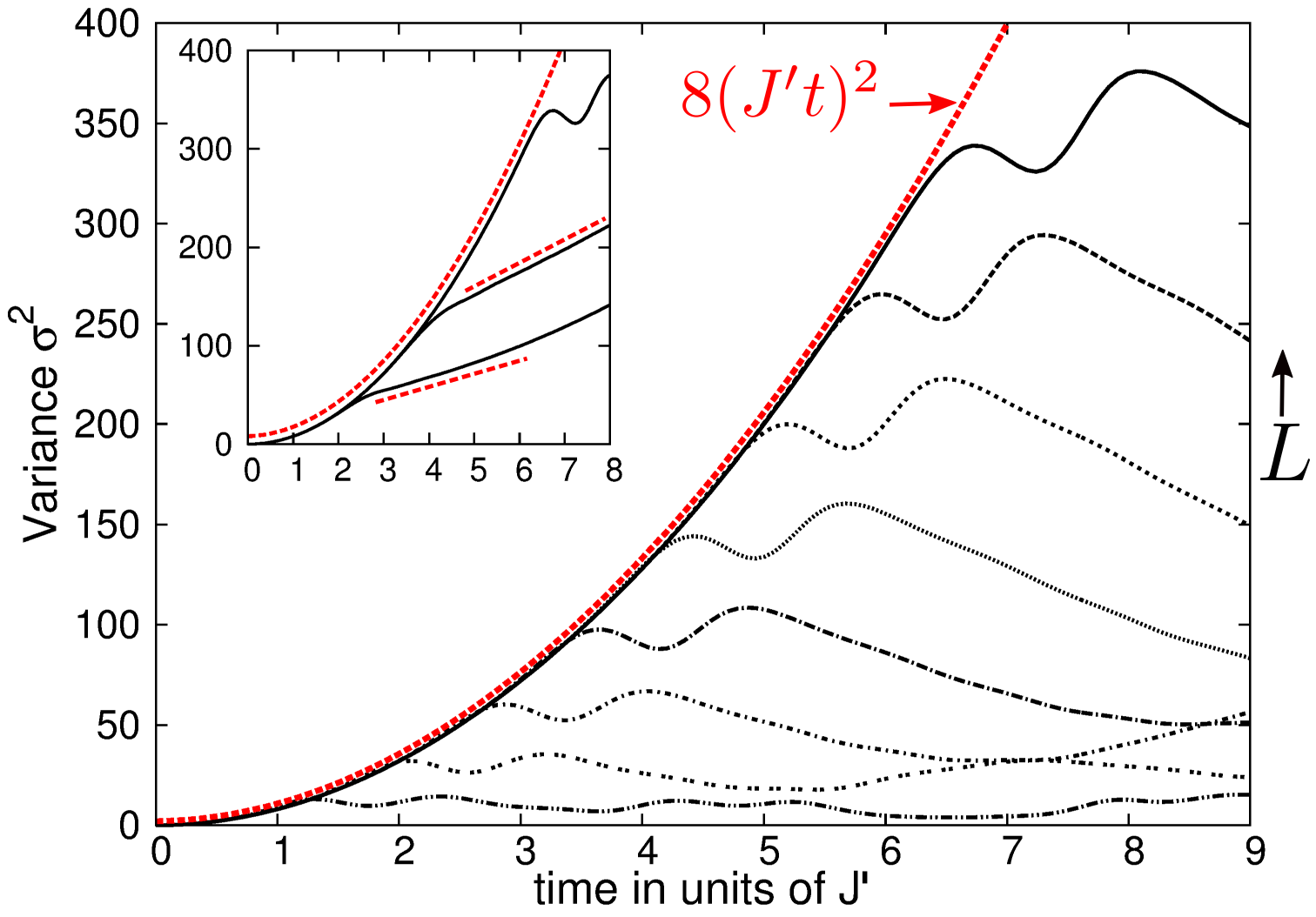}}
\caption{\label{fig:6}(Color online) Variance $\sigma^2$ of the dipole wavefunction for
 $L=\{12,18,\cdots,54\}$ from down to up. The red-dasehd line represent is the theoretical
prediction in Eq.~(\ref{eqn:18}). The inset shows the dependence of $\sigma^2$ on the 
location $l_0$ of the initial dipole $l_0=27$ at the center of the lattice and 
$l_0=\{8,14\}$ (The two lowest ones) for a fixed lattice sites number $L=54$.
The straight red-dashed lines correspond to the normal diffusion tendency of the wavefunction
spreading right after its contact with the boundaries.}
\end{figure}
The dipole wavefunction then diffuses anomalously since $\sigma^2\propto t^2$. The spreading of the wavefuction 
is then faster than the one for Brownian particles or a particle in a classical random walk. This behavior is exhibited 
in Fig.~(\ref{fig:6}) where we have computed the variance of the dipole wavefunction as a function of time 
for increasing the system size $L$ where, as shown above, before arriving to the first edge the dipole wavefunction 
undergoes the ballistic spreading. 

Boundary effects on the dyamics are usually much complicated to be analitically computed. We numerically
study the border effect by initializing the wavefunction at different location in the lattice, that is, 
as a function of $l_0$. We set open boundary conditions. The results are plotted in the inset of Fig.~(\ref{fig:6}) 
for a lattice with $L=54$. The dynamical variance for the first edge-to-edge transit preserves the power-law dependence 
obtained in Eq.~(\ref{eqn:18}), while for backwards transit right after the reflection at the boundary
the diffusive spreading slows down. The plot is done for the initial locations $l_0=27$, i.e., right in the center
of the lattice, and two cases closer to the border for which $l_0=\{8,14\}$. The dipole wavefunction 
initialized close to the border undergoes a backwards diffusive transit after reflecting at the borders  
for which its variance now changes linearly in time, $\sigma^2\propto t$ (see red-dashed straight lines in the inset). 
It is then expected that for a dipole set initially at one of the border the spreading of the wavefunction 
becomes nearly normally diffusive in its first edge-to-edge transit. 

\subsection{Long-time dynamics}
The self-interference pattern in Fig.~(\ref{fig:2}) is a finite size effect and appears no matter 
which of the two classes of boundary conditions we imposed, i.e., the open or the periodic ones. 
However, relocalization of the dipole wavefunction is seen to be only possible for open boundary conditions 
which admit recurrences for finite lattice size. We have also seen that the delocalization-localization 
process is periodic (see Fig.~(\ref{fig:2}-\ref{fig:4})) in time, whose period $T$ can be extracted from 
the IPR~(\ref{eqn:10}) and analyzed in terms of the system size.
\begin{figure}[t!]
\centering{\includegraphics[width=0.8\columnwidth]{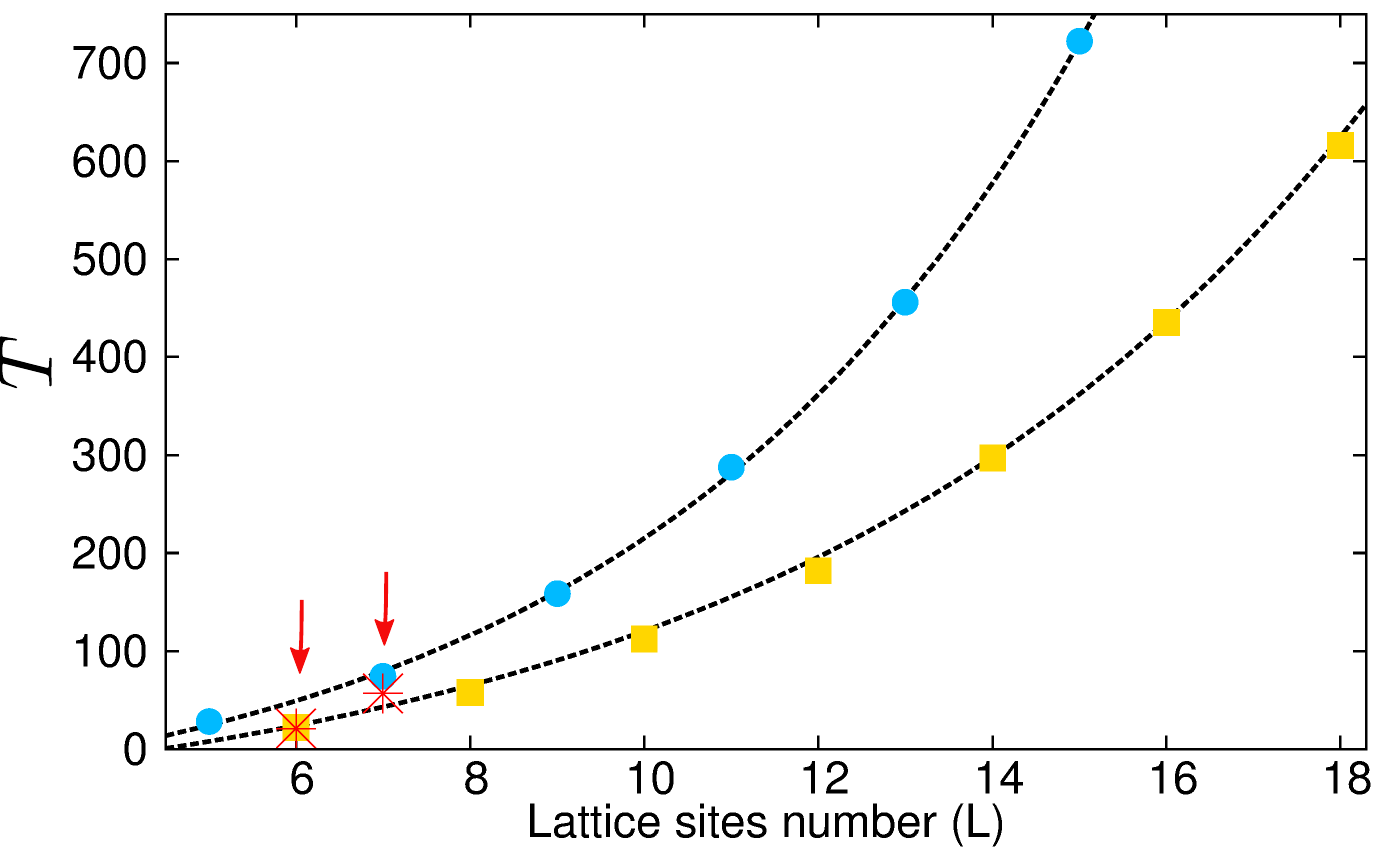}}
\caption{\label{fig:7}(Color online) Interference time as a function of system size $L$, with $L=\{5,6,\cdots,18\}$,
for odd $L$ ($\bullet$) and even ($\blacksquare$). The lines represent the best fits and the red asterisk are the interference 
time computed with the full model for lattices with $L=6$ and $L=7$ sites. These calculations were
done for open boundary conditions.} 
\end{figure}
\begin{figure}[t!]
\centering{\includegraphics[width=0.95\columnwidth]{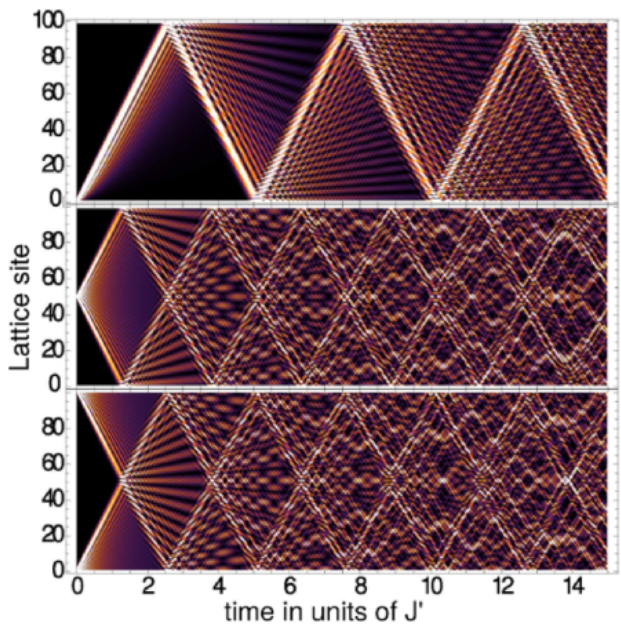}}
\caption{\label{fig:8}(Color online) Density profile of the dipole wavefunction propagation along the lattice for a chain with
$L=100$ sites. The initial condition corresponds to (Top) a dipole placed in one corner $|0100\cdots 000\rangle$,
(Middle) a dipole in the middle of the lattice $|00\cdots 010\cdots 00\rangle$, both cases with open boundary conditions,
and (Bottom) periodic boundary conditions and the same initial state as in the top panel.}
\end{figure}

Fig.~\ref{fig:7} shows the dependence of $T$ on the increasing lattice site number $L$ and on its parity.
For $L$ odd the self-interference pattern is preserved longer than for the case of $L$ even. In both cases 
$T$ grows exponentially with the system size. This result, valid only for open boundary conditions, cannot 
be extrapolated to a thermodynamic limit $L\rightarrow\infty$ because the self-interference occurs as a 
consequence of the reflecting borders. For periodic boundary conditions, the interference appears because
of the dynamical wrapping of the wavefunction on the surface of a cylinder of diameter $L/\pi$. Thus, 
the interference pattern is a consequence of the encounter of counterpropagating principal  
and secondary waves (see Fig.~\ref{fig:8}), an effect that has not recurrences. 
Therefore $T\rightarrow \infty$ presenting an unique and everlasting interference pattern.
The complex dynamical pattern observed in Fig.~\ref{fig:8} is nothing but the so-called quantum 
carpet ~\cite{Berry2001}. 

Self-interference might be understood by inspecting the delocalization process of the dipole wavefunction 
when travelling along the lattice. The transit from one site to the nearest neighbor induces, 
due to local reflections, small secondary waves-like components which also travel in the same direction 
of the principal wave when the dipole is initialized at one of the edges. This is also clear after 
inspecting Eq.~(\ref{eqn:15}) within the time regime for $0<t\ll 1/4J'$. Yet, those secondary waves are 
retarded in time arriving late at the edge interfering with the already reflected principal wave 
(see Fig.~\ref{fig:8}-(middle)). This generates a final and sufficiently complex interference pattern, 
the quantum carpet. In the case of a dipole initialized, for example, in the middle of the lattice, 
two principal waves appears owing to the translational invariance
of our system. Thus, the generated secondary waves travel both directions then triggering intereference
effect at both edges of the lattice simultaneously (see Fig.~\ref{fig:8}-(middle)). The same occurs for an 
initial dipole state at one of the edges when imposing with periodic boundary conditions for which also 
intereference effects appear in the middle of the lattice (see Fig.~\ref{fig:8}-(bottom)).\\

\section{SUMMARY AND CONCLUSIONS}
\label{sec:5}
We have studied a single-band tilted Bose-Hubbard Hamiltonian in one dimension showing that far from 
single particle and two-particle level resonances there is still very interesting dynamical effects.
The phenomenology induced by the propagation of a quasiparticle made of a doublon and a holon, that we
called dipole excitation or exciton, shows the appearance of finite size dynamical self-interference 
pattern. This effect is known as a quantum carpet and can be understood as the interference of the principal and
secondary wave components of the time-evolved dipole wavefunction. We also show that for large
but finite system sizes $L$ the emergent interference pattern lasts very long, and its caracteristic time
$T$ grows exponentially with $L$. We derive an analytic expression for the dipole site occupation for short times
that allow us to determine the diffusive properties of our system. Here, it has been shown that the spreading  
of the wavefunction is superdiffusive, a ballistic behavior similar to the one presented in 
Refs.~\cite{Andraschko2015,Hofmann2012}, yet for a doublon-holon excitation in a unit-filled tilted optical
lattice, and not in an empty one. Furthermore, we study the effect of the borders in the dynamics showing that right 
after the border, for a dipole initialized close to the edges, the diffusive spreading is slowed down becoming 
normal, that is, the wavefunction variance grows linearly with the time.

The results presented in this paper can be straightforwardly implemented in experiments using ultracold atoms 
trapped in optical lattices as, for instance, an extension of the works done by F. Meinert {\it et al}
\cite{Meinert2013,Meinert2014} and Simon {\it et al.} \cite{Simon2011}. We suggest to explore the regime of 
strong tilts, Stark fields, far from the many-particle first order of resonances where our results 
take place.\\

\section{Acknowledgments}
The authors acknownledge the finantial support of the University del Valle (project CI 7996).
C. A. Parra-Murillo greatfully acknowledges the financial support of COLCIENCIAS (grant 656). 
We warmly thank to Dr. A. Arg\"uelles for the lively discussion and critical reading of our manuscript. 

\end{document}